\documentclass[10pt,twocolumn,letterpaper]{article}

\usepackage{cvpr}              

\usepackage[most]{tcolorbox}

\newtcolorbox{disclaimerbox}{
    colback=gray!10,     
    colframe=gray!40,    
    boxrule=0.5pt,       
    arc=4pt,             
    auto outer arc,
    boxsep=5pt,
    left=6pt,
    right=6pt,
    top=4pt,
    bottom=4pt,
    enhanced jigsaw
}

\usepackage{lipsum}
\usepackage{comment}
\usepackage{svg}
\usepackage{tikz}
\usepackage{colortbl}
\usetikzlibrary{positioning, arrows.meta, chains, calc}

\definecolor{cvprblue}{rgb}{0.21,0.49,0.74}
\definecolor{lblue}{rgb}{0.9,0.95,1}
\definecolor{lpurple}{rgb}{0.35,0.25,0.55}
\definecolor{lgreen}{rgb}{0.95,1,0.95}
\definecolor{sblue}{rgb}{0,0.45,1}
\definecolor{lgray}{gray}{0.95}
\definecolor{lyellow}{rgb}{1,1,0.92}

\usepackage[pagebackref,breaklinks,colorlinks,citecolor=cvprblue]{hyperref}

\def\paperID{*****} 
\def\confName{CVPR}
\def\confYear{2025}

\title{RAW Image Reconstruction from RGB on Smartphones.\\NTIRE 2025 Challenge Report}

\author{
Marcos V. Conde~$^{*\dagger}$ \and
Radu Timofte~$^{*}$ \and
Radu Berdan \and Beril Besbinar \and Daisuke Iso \and
Pengzhou Ji \and Xiong Dun \and Zeying Fan \and
Chen Wu \and Zhansheng Wang \and Pengbo Zhang \and Jiazi Huang \and
Qinglin Liu \and Wei Yu \and Shengping Zhang \and Xiangyang Ji \and
Kyungsik Kim \and Minkyung Kim \and Hwalmin Lee \and
Hekun Ma \and Huan Zheng \and Yanyan Wei \and Zhao Zhang \and
Jing Fang \and Meilin Gao \and Xiang Yu \and
Shangbin Xie \and Mengyuan Sun \and Huanjing Yue \and Jingyu Yang
Huize Cheng \and Shaomeng Zhang \and Zhaoyang Zhang \and Haoxiang Liang
}

\begin{document}

\maketitle

\let\thefootnote\relax\footnotetext{$*$ Marcos V. Conde ($\dagger$ corresponding author, project lead) and Radu Timofte are the challenge organizers, while the other authors participated in the challenge and survey. \\
$*$~University of W\"urzburg, CAIDAS \& IFI, Computer Vision Lab.\\ 
NTIRE 2025 webpage:~\url{https://cvlai.net/ntire/2025}.\\
Code:~\url{https://github.com/mv-lab/AISP}} 

\begin{abstract}
Numerous low-level vision tasks operate in the RAW domain due to its linear properties, bit depth, and sensor designs. Despite this, RAW image datasets are scarce and more expensive to collect than the already large and public sRGB datasets. For this reason, many approaches try to generate realistic RAW images using sensor information and sRGB images. This paper covers the second challenge on RAW Reconstruction from sRGB (Reverse ISP). We aim to recover RAW sensor images from smartphones given the corresponding sRGB images without metadata and, by doing this, ``reverse" the ISP transformation. Over 150 participants joined this NTIRE 2025 challenge and submitted efficient models. The proposed methods and benchmark establish the state-of-the-art for generating realistic RAW data.
\end{abstract}

\setlength{\abovedisplayskip}{1pt}
\setlength{\belowdisplayskip}{1pt}

\section{Introduction}
Most low-level vision and computational photography tasks heavily rely on RGB images produced by the camera built-in Image Signal Processor (ISP)~\cite{conde2022model, delbracio2021mobile, karaimer2016software}. The ISPs convert RAW sensor data into visually appealing RGB images tailored to human perception. The widespread availability of RGB datasets has significantly accelerated research into modeling the RAW-to-RGB transformation using deep neural networks \ie learned ISPs~\cite{ schwartz2018deepisp, ignatov2020replacing, ignatov2022learned, conde2022model}.

Nonetheless, RAW sensor data inherently offers unique advantages due to its linear relationship with scene irradiance, higher bit depth (typically 12–14 bits), and preservation of unaltered sensor noise. These attributes make RAW data particularly beneficial for tackling inverse problems common in low-level vision, such as image denoising, deblurring, and super-resolution~\cite{qian2019trinity, gharbi2016deep, Liu2020joint, conde2024bsraw, conde2024rawir}. Furthermore, professional photographers frequently prefer processing RAW images manually to achieve greater control and superior visual quality~\cite{karaimer2016software}.

However, the limited availability and diversity of RAW image datasets severely constrain the potential of deep learning approaches. To address this limitation, several methods have been proposed to reconstruct realistic RAW data from widely accessible RGB images. Some approaches assume a model-based ISP and use metadata (\ie white balance gains, color correction matrices) to reconstruct the RAW images~\cite{Nguyen_2016_CVPR, brooks2019unprocessing, punnappurath2021spatially, Seo_2023_ICCVgraphicsraw} utilize camera-specific metadata to reverse the ISP process. While effective, these approaches incur practical overheads by requiring additional metadata storage. Moreover, metadata (and ISP parameters information) is rarely available.

Recent advancements in learning-based strategies aim to eliminate dependence on metadata and prior information about the ISP by learning the RAW reconstruction directly from RGB images~\cite{zamir2020cycleisp, xing2021invertible,conde2022model, dong2022rispnet, berdan2025reraw}. These techniques have demonstrated promising results by learning mappings between RAW and RGB domains. 

Motivated by recent developments, we introduce the NTIRE 2025 RGB-to-RAW Challenge, based on the first edition ``Reversed Image Signal Processing and RAW Reconstruction"~\cite{conde2022reversed}. The challenge focuses on advancing methods for realistic RAW reconstruction directly from smartphone RGB images without relying on metadata. With over 150 participants contributing, this challenge significantly advances the state-of-the-art in RAW reconstruction.


\section{NTIRE 2025 RGB-to-RAW Challenge}
\label{sec:challenge}

\subsection{Dataset}

We propose a novel dataset for this challenge using diverse \textbf{smartphones}. Unlike previous datasets employed for this task~\cite{berdan2025reraw}, we use smartphones instead of DSLR and DSLM images since their ISPs are considered more complex~\cite{delbracio2021mobile}, thus, recovering the RAW images is harder. Moreover, the degradations present in smartphone images are more notable than in DSLR and DSLM cameras.

The RAW-RGB pairs are manually filtered to ensure diversity and natural properties (\ie remove extremely dark or overexposed images). The dataset includes images with different levels of noise and illumination, including day and night images.

We use the following camera devices: iPhone X (Sony Exmor RS), Samsung S9 (Sony IMX345), Samsung S21 (Sony IMX616 Quad-Bayer sensor) and Vivo X90 (Sony IMX866). The dataset \textbf{pre-processing} is as follows: 

\begin{itemize}
    \item All the RAW images in this dataset have been standardize to follow a Bayer Pattern RGGB, and already white-black level corrected.

    \item  Each RAW image was split into several crops of size $512 \times 512 \times 4$ ($1024 \times 1024 \times 3$ for the corresponding RGBs). For each RAW-RGB pair we provide to the participants the corresponding metadata including color correction matrices, white balance gains and other useful ISP parameters. For the test images, there is \emph{no explicit metadata} \ie participants might infer the ISP parameters.
    
    \item The RGB images are the corresponding captures from the phone \ie the phone imaging pipeline (ISP) output. We do not render the RGB images using simple software such as \texttt{rawpy}.

    \item The dataset is publicly available at \url{https://huggingface.co/datasets/marcosv/rgb2raw}
\end{itemize}

\paragraph{Training} We provide the participants $1024 \times 1024 \times 4$ clean high-resolution (HR) RAW images. The training set includes only images from the iPhone X (972 pairs) and Samsung S9 (474 pairs) -- a filtered set from the RAW2RAW dataset~\cite{afifi2021raw2raw}. During training, participants can use the ISP metadata to train and fine-tune their models.

\paragraph{Testing} The test set includes images captured using the training (target) devices, and unknown (OOF) devices such as Samsung S21 and Vivo X90, which also represent more modern sensors. Thus, we want to test the methods ability to recover RAW images from known and unknown sensors, even considering design gaps. During testing, the participants do not have access to the reference RAW images and ISP metadata. The target device test set contains 120 images, while the OOF test set contains 60 images.

\subsection{Baselines}
Since metadata is not available during testing, we use as baseline pure deep learning-based approaches. ReRAW~\cite{berdan2025reraw} represents the \emph{state-of-the-art} on RAW image reconstruction for DSLR and DSLM cameras. Also, DualRAW (see Sec.~\ref{sec:dualraw}) represents an advanced neural network for RAW image processing and reconstruction. 

\subsection{Results}

In \cref{tab:rawre-benchmark} we provide the challenge benchmark. We calculate the PSNR and SSIM metrics on uncompressed 12-bit RAW images. We separate methods in two tracks: efficient and general -- efficient methods are limited to 0.2M parameters. Many methods achieve high fidelity metrics on the known devices, while only the ``simple" and efficient methods avoid overfitting and generalize on unknown OOF devices. We highlight DBNet~(see Sec.~\ref{sec:tongji}) as the best proposed method. We provide qualitative results in the challenge repository~\footnote{\url{https://github.com/mv-lab/AISP/}}. Moreover, we summarize the technical details of the proposed methods in Table~\ref{tab:summary}, including number of parameters.

\begin{table*}[t]
    \centering
    \resizebox{0.85\textwidth}{!}{%
    \begin{tabular}{rccccccc}
        \toprule
        \textbf{Method} & \multicolumn{2}{c}{\textbf{Overall}} & \multicolumn{2}{c}{\textbf{Target Devices}} & \multicolumn{2}{c}{\textbf{OOF Devices}} & \textbf{Track} \\
        \cmidrule(lr){2-3} \cmidrule(lr){4-5} \cmidrule(lr){6-7}
        & PSNR & SSIM & PSNR & SSIM & PSNR & SSIM & \\
        \midrule
        \rowcolor{lgray} DualRAW~\ref{sec:dualraw}   & 26.50 & 0.7537 & 29.49 & 0.8274 & 22.93 & 0.6653 & general \\
        \rowcolor{lgray} ReRAW~\cite{berdan2025reraw}     & 24.52 & 0.6988 & 26.90 & 0.7820 & 21.66 & 0.5989 & general \\
        GAR2Net~\ref{sec:ivislab}   & 26.98 & 0.7399 & 31.22 & 0.8694 & 21.89 & 0.5844 & general \\
        
        TDMFNet~\ref{iirlab}    & 26.30 & 0.7150 & 28.02 & 0.8097 & 21.15 & 0.6030 & general 
        \\
        ResUNet~\ref{sec:lvg}   & 24.54 & 0.6981 & 24.01 & 0.6803 & 25.17 & 0.7196 & general \\
        VIP~\ref{sec:vip}       & 26.99 & 0.7543 & 31.77 & 0.8762 & 21.26 & 0.6080 & general \\
        UNAFNet~\ref{sec:unafnet}   & 26.87 & 0.7608 & 29.93 & 0.8436 & 23.20 & 0.6615 & general \\
        ULite~\ref{sec:unisoc}    & 26.31 & 0.7653 & 29.49 & 0.8406 & 22.49 & 0.6750 & general \\
        
        \midrule
        \rowcolor{lyellow} 
        DBNet~\ref{sec:tongji} & 27.66 & 0.7700 & 30.76 & 0.8353 & 23.94 & 0.6916 & efficient \\
        
        ULite~\ref{sec:unisoc}    & 26.11 & 0.7621 & 29.41 & 0.8416 & 22.15 & 0.6666 & efficient \\
        
        GAR2Net~\ref{sec:ivislab}    & 25.02 & 0.7181 & 28.58 & 0.8162 & 20.74 & 0.6004 & efficient \\
        
        \bottomrule
    \end{tabular}
    }
    \caption{\textbf{NTIRE 2025 RAW Image Reconstruction from RGB on Smartphones.} We provide the \textbf{SSIM/PSNR} results on the testing set. All the metrics are calculated in the RAW domain. We highlight in gray the baseline methods. The efficient track includes models under 0.2M parameters and able to process 12MP images. The simple models generalize better on unknown OOF devices.}
    \label{tab:rawre-benchmark}
\end{table*}

\subsection*{Related Computer Vision Challenges}

This challenge is one of the NTIRE 2025 \footnote{\url{https://www.cvlai.net/ntire/2025/}} Workshop associated challenges on: 
ambient lighting normalization~\cite{ntire2025ambient}, 
reflection removal in the wild~\cite{ntire2025reflection}, 
shadow removal~\cite{ntire2025shadow}, 
event-based image deblurring~\cite{ntire2025event}, 
image denoising~\cite{ntire2025denoising}, 
XGC quality assessment~\cite{ntire2025xgc}, 
UGC video enhancement~\cite{ntire2025ugc}, 
night photography rendering~\cite{ntire2025night}, 
image super-resolution (x4)~\cite{ntire2025srx4},
real-world face restoration~\cite{ntire2025face}, 
efficient super-resolution~\cite{ntire2025esr}, 
HR depth estimation~\cite{ntire2025hrdepth}, 
efficient burst HDR and restoration~\cite{ntire2025ebhdr}, 
cross-domain few-shot object detection~\cite{ntire2025cross}, 
short-form UGC video quality assessment and enhancement~\cite{ntire2025shortugc,ntire2025shortugc_data}, 
text to image generation model quality assessment~\cite{ntire2025text}, 
day and night raindrop removal for dual-focused images~\cite{ntire2025day}, 
video quality assessment for video conferencing~\cite{ntire2025vqe}, 
low light image enhancement~\cite{ntire2025lowlight}, 
light field super-resolution~\cite{ntire2025lightfield}, 
restore any image model (RAIM) in the wild~\cite{ntire2025raim}, 
raw restoration and super-resolution~\cite{ntire2025raw}, 
and raw reconstruction from RGB on smartphones~\cite{ntire2025rawrgb}.

\subsection*{Acknowledgments}
This work was partially supported by the Humboldt Foundation. We thank the NTIRE 2025 sponsors: ByteDance, Meituan, Kuaishou, and University of Wurzburg (Computer Vision Lab). The challenge organizers appreciate the discussions and expert advise from Radu Berdan, Beril Besbinar, and Daisuke Iso (Sony AI).

\section{Challenge Methods for RGB-to-RAW}
\label{sec:teams}

\vspace{2mm}

\begin{disclaimerbox}
In the following Sections we describe the top challenge submissions. Note that the method descriptions were provided by each team as their contribution to this report.
\end{disclaimerbox}

\subsection{DualRAW - Dual Intenstiy sRGB to RAW Reconstruction}
\label{sec:dualraw}

\begin{center}

\vspace{2mm}
\noindent\emph{\textbf{Sony AI}}
\vspace{2mm}

\noindent\emph{Beril Besbinar, Daisuke Iso}





\end{center}


DualRAW draws significant inspiration from RawHDR~\cite{zou2023rawhdr}, which reconstructs HDR images from RAW sensor data using exposure masks and dual intensity guidance.
RawHDR's design stems from the observation that green spectral integration and channel averages exceed those of red and blue.
Consequently, they propose that red and blue channels lose detail in low-light regions during RAW-to-HDR mapping, while green channels are more prone to detail loss in highlights.

This idea and RawHDR's design helped us create DualRAW.
Since sRGB images are optimized for human perception, which is more sensitive to changes in darker tones than in brighter ones, we think reconstructing green channels differently from red and blue could lead to easier optimization of the proposed learning method.
For this, we use two encoders, $f_{\text{enc}}^O$ and $f_{\text{enc}}^U$ to process the input sRGB image, $\mathbf{X}_{\text{RGB}}$.
We apply a de-gamma operation to $\mathbf{X}_{\text{RGB}}$ to give the encoders two different versions of the input.
The encoders produce feature maps, $\mathbf{Y}_{\text{over}}$ and $\mathbf{Y}_{\text{under}}$.
We also use mask estimation modules $f_{\text{m}}^O$ and $f_{\text{m}}^U$ to estimate over- and under-exposure masks, $\mathbf{M}_{\text{over}}$ and $\mathbf{M}_{\text{under}}$, respectively.
\begin{align}
    \vspace{-1cm}
     \mathbf{Y}_{\text{over}} &= f_{\text{enc}}^O(\mathbf{X}_{\text{RGB}}, \mathbf{X}_{\text{RGB}}^\gamma) \\
     \mathbf{Y}_{\text{under}} &= f_{\text{enc}}^U(\mathbf{X}_{\text{RGB}}, \mathbf{X}_{\text{RGB}}^\gamma) \\
     \mathbf{M}_{\text{over}} &= f_{\text{m}}^O(\mathbf{X}_{\text{RGB}}) \\
     \mathbf{M}_{\text{under}} &= f_{\text{m}}^U(\mathbf{X}_{\text{RGB}})
\end{align}

The main image representation is a weighted combination of $\mathbf{Y}_{\text{over}}$ and $\mathbf{Y}_{\text{under}}$ feature maps combined with a global context $\mathbf{Y}_{\text{global}}$ from a global encoder $f_{\text{global}}$.
This combined representation is then fed to the reconstruction module, $f_{\text{rec}}^{\text{RGGB}}$ that outputs a 4-channel image.
The over- and under-exposed feature pathways give us residual outputs, $\mathbf{X}_{\text{RAW}}^{\text{BB}}$ and $\mathbb{X}_{\text{RAW}}^{\text{GG}}$, respectively, to account for the differences in the red-blue and green channel properties.
\begin{align}
    \mathbf{X}_{\text{RAW}}^{\text{RB}} &= f_{\text{rec}}^\text{RB}(\mathbf{Y}_{\text{under}}) \\
    \mathbf{X}_{\text{RAW}}^{\text{GG}} &= f_{\text{rec}}^\text{GG}(\mathbf{Y}_{\text{over}}) \\
    \mathbf{Y} &= \mathbf{Y}_{\text{global}} + \mathbf{M}_{\text{under}} \odot \mathbf{Y}_{\text{under}} +  \mathbf{M}_{\text{over}} \odot \mathbf{Y}_{\text{over}} \\
    \mathbf{X}_{\text{RAW}}^{\text{RGGB}} &= f_{\text{rec}}^\text{RGGB}(\mathbf{Y}) +  \mathbf{X}_{\text{RAW}}^{\text{RB}} + \mathbf{X}_{\text{RAW}}^{\text{GG}}
\end{align}

\begin{figure*}[t]
    \centering
    \includegraphics[width=\textwidth]{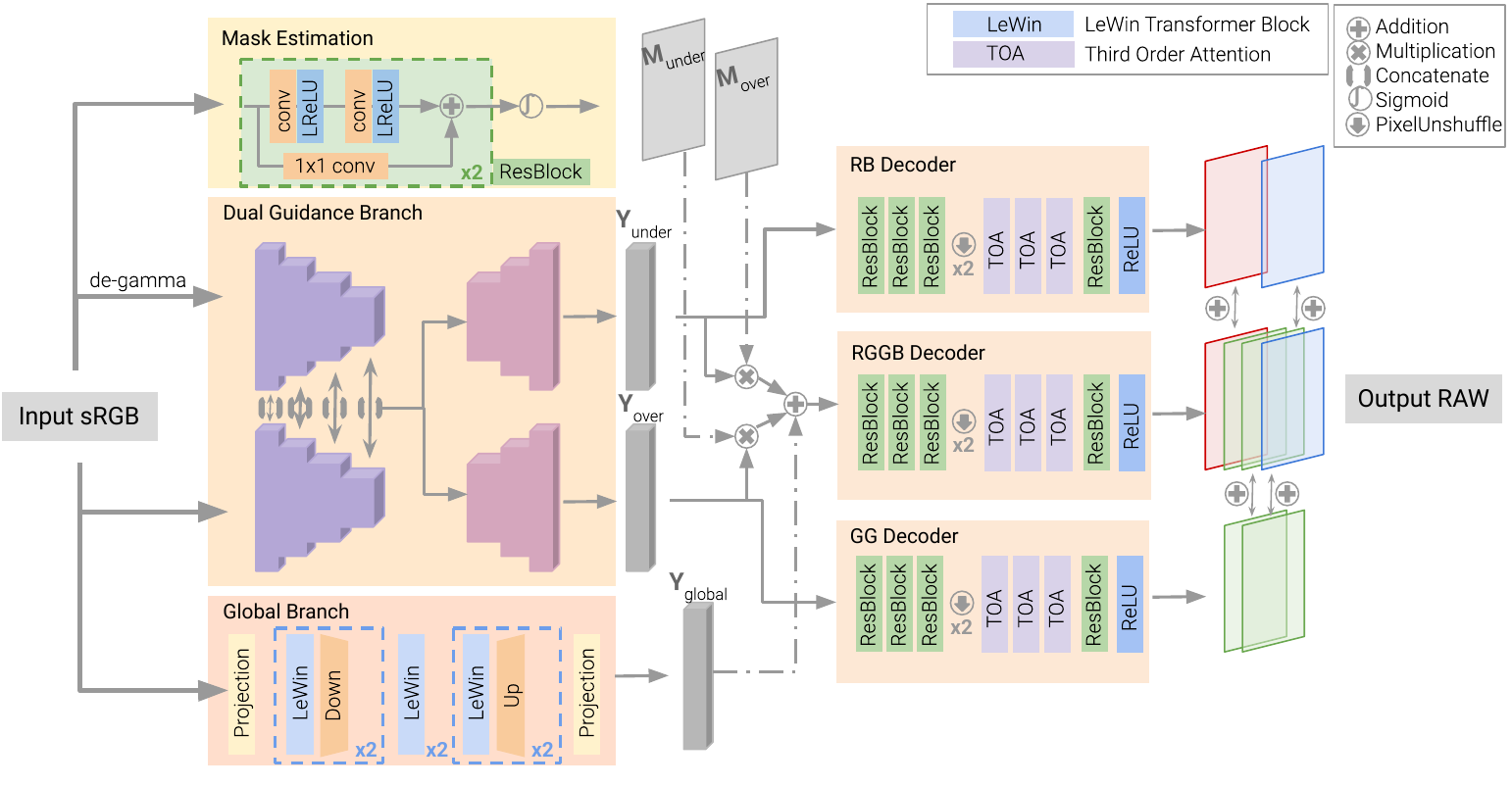}
    \caption{DualRAW Architecture: A Uformer-based Global Branch extracts global features from the input RGB image. Parallel UNET encoders process the RGB image, with one applying a de-gamma operation. Encoder features are concatenated and fed to two UNET decoders, generating over- and under-exposed embeddings. These embeddings are weighted using exposure masks predicted by simple convolutional networks with residual blocks. A final decoder reconstructs the output RAW image, employing separate residual connections for red-blue and green channels to account for distinct histogram characteristics.
    \vspace{2mm}
    }
    \label{fig:dualraw_arch}
\end{figure*}

An illustration of the proposed pipeline is presented in Figure \ref{fig:dualraw_arch}.

\paragraph{Implementation details}
Our implementation also mainly follows RawHDR.
The feature encoding functions, $f_{\text{enc}}^{(.)}$, resemble UNET~\cite{ronneberger2015u} with $2^4$ times downsampling on the contracting path, where the feature map at the highest spatial resolution has 32-channels.
On the other hand, mask estimation modules $f_{\text{m}}^{(.)}$  are implemented as simple convolutional networks with two residual blocks.
A final sigmoid activation ensures the value range of the estimated masks.
The global context encoder is a U-shaped image transformer~\cite{wang2022uformer}.
Finally, the image reconstruction networks $f_{\text{rec}}^\text{(.)}$ are composed of three residual blocks, followed by a pixel unshuffling operation~\cite{shi2016real} and three blocks of Third Order Attention (TOA)~\cite{dong2022rispnet}.

For training DualRAW model, we use a combination of log-L2 loss, $\mathcal{L}_{\log\text{L2}}$, clipped L1 loss~\cite{zhu2020eednet},  $\mathcal{L}_{\text{clippedL1}}$, mask loss $\mathcal{L}_{\text{mask}}$~\cite{zou2023rawhdr} and LPIPS loss~\cite{zhang2018unreasonable} $\mathcal{L}_{\text{LPIPS}}$ with $\tau_1 = 0.2$ and $\tau_2 = 0.5$ in Equation \ref{eq:loss}.

\begin{equation}
    \mathcal{L} = \mathcal{L}_{\log\text{L2}} + \mathcal{L}_{\text{clippedL1}} + \tau_1 \mathcal{L}_{\text{mask}} + \tau_2 \mathcal{L}_{\text{LPIPS}}
    \label{eq:loss}
\end{equation}

The overall pipeline is implemented and trained in PyTorch.
DualRAW model is trained with AdamW~\cite{loshchilov2017decoupled} optimizer for 200 epochs using a triangular cyclic learning rate~\cite{smith2017cyclical} using only the training dataset provided by the NTIRE Challenge with an effective batch size of 8.
Images are used at their full resolution, $1024\times1024$ to ensure the context encoder captures the most relevant information.
Only horizontal and vertical flipping are used for data augmentation.

\begin{table}[h!]
    \centering
    \begin{tabular}{c c}
         Input Size & Inference Time  \\
         \hline
         $1024 \times 1024$   & 87ms  \\
         $3072 \times 2048$   & 519ms  \\
         $4096 \times 3072$   & 1.018s  \\
    \end{tabular}
    \caption{The inference time of DualRAW with varying input sizes on a single Nvidia H100 GPU}
    \label{tab:inference_times}
\end{table}

The model has 1.6M trainable parameters and inference times for inputs of variant sizes could be found in Table \ref{tab:inference_times}.

\subsection{ReRAW: RGB-to-RAW Reconstruction via Stratified Sampling}
\label{sec:reraw}

\begin{center}

\vspace{2mm}
\noindent\emph{\textbf{Sony AI}}
\vspace{2mm}

\noindent\emph{Radu Berdan, Daisuke Iso}





\end{center}


ReRAW~\cite{berdan2025reraw} is designed to reconstruct a $W/2 \times H/2 \times 4$ packed RGGB (RAW) image given a $W \times H \times3$ RGB image. As a difference from the original paper, in this implementation ReRAW can handle direct high resolution image re-construction in a single pass. Alternatively, it can also be convolved over an input RGB image to reconstruct the full required RAW image patch-by-patch, if resources are limited.

The model starts by encoding general characteristics from the original RGB image (scaled to $128 \times 128$) such as luminosity and color space features, and uses this infomation to modulate the RGB-to-RAW color conversion.

The model then uses a multi-head architecture to predict raw patches in gamma space, over multiple gamma candidates. Gamma-corrected patch candidates are re-linearized (by applying an inverse gamma process) and proportionally averaged by a weight vector predicted by a Gamma Scaling Encoder from the original full RGB image. In this way, the model learns to select input  image-dependent gamma trans-formations that would facilitate a better RAW conversion. Additionally, training via a stratified sampling data selection technique helps in mitigating the skew of pixel values commonly found in RAW images.

\paragraph{Implementation Details}

Training data consists of a mix of the provided training set from the challenge organizers, both clean and noisy picture pairs, as well as a subset of the FIVEK dataset (pairs from the Nikon and Canon cameras).

Training data preparation involves performing stratified sampling as described in the original paper\cite{berdan2025reraw}. About 30k of paired $66\times66$ RGB and $32\times32$ RAW patches have been sampled from the combined datasets. Training was performed using a batch size of 32, over 128 epochs, under a cosine annealing schedule with warm restarts every 16 epochs, starting learning rate of 1e-3 decaying to 1e-5, and ADAM optimizer. More details in the original paper~\cite{berdan2025reraw}.

\begin{figure*}
    \centering
    \includegraphics[width=\linewidth]{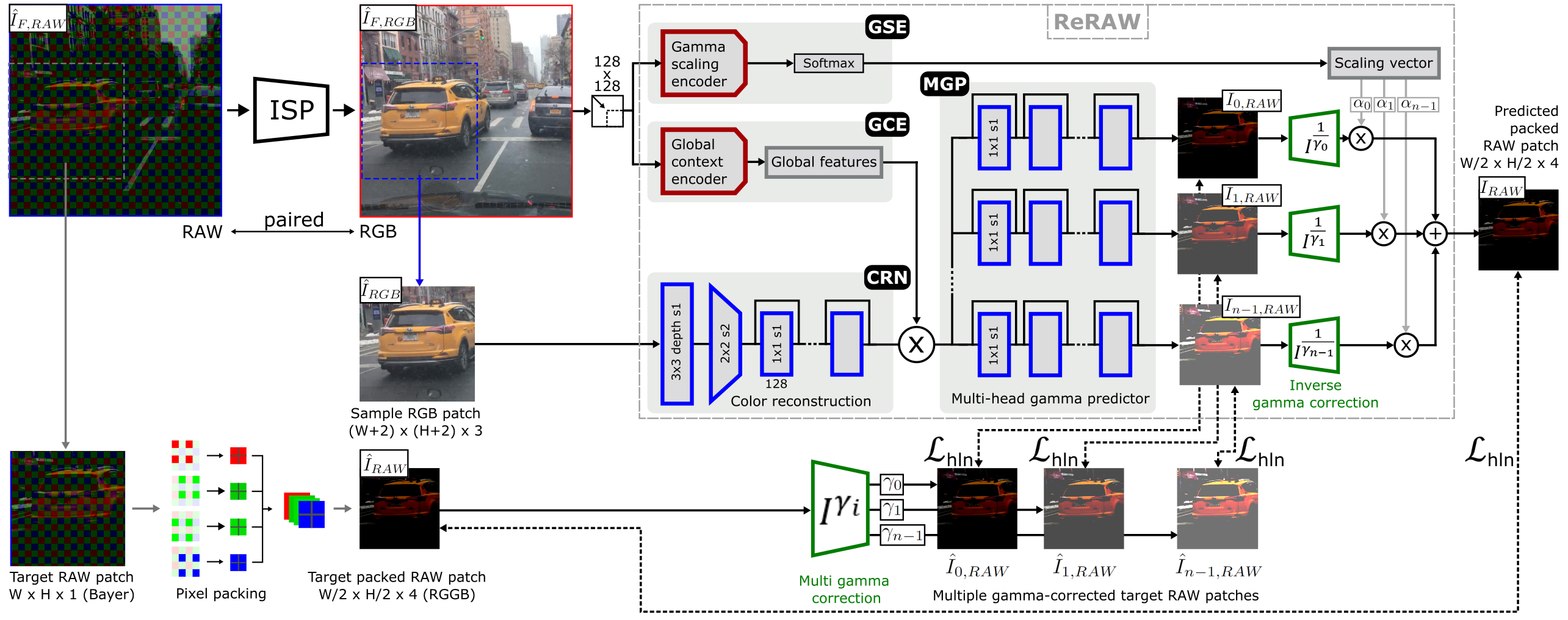}
    \caption{Illustration of ReRAW~\cite{berdan2025reraw} architecture and training data flow. A Global Context Encoder (GCE) extracts features from the full RGB image to guide the Color Reconstruction network (CRN), while a Multi-head Gamma Predictor (MGP) generates multiple gamma-corrected RAW patches. These patches are then de-gammaed (inverse gamma correction), scaled by a scaling vector, predicted by a Gamma Scaling Encoder (GSE) from the original RGB image, and summed to form the final RAW patch. Losses are applied between each intermediate gamma-corrected RAW patch and target, as well as between the final RAW output and target RAW.
    \vspace{2mm}
    }
\label{fig:reraw}
\end{figure*}
\subsection{DBNet: A dynamical bias convolution network}
\label{sec:tongji}

\begin{center}

\vspace{2mm}
\noindent\emph{\textbf{Team TongJi-IPOE}}
\vspace{2mm}

\noindent\emph{Pengzhou Ji~$^1$,
Xiong Dun~$^1$,
Zeying Fan~$^1$,}

\vspace{2mm}

\noindent\emph{Institute of Precision Optical Engineering, School of Physics Science and Engineering, Tongji University}

\vspace{2mm}


\end{center}


\paragraph{Method Description}

\begin{figure}[t]
    \centering
    \includegraphics[width=\linewidth]{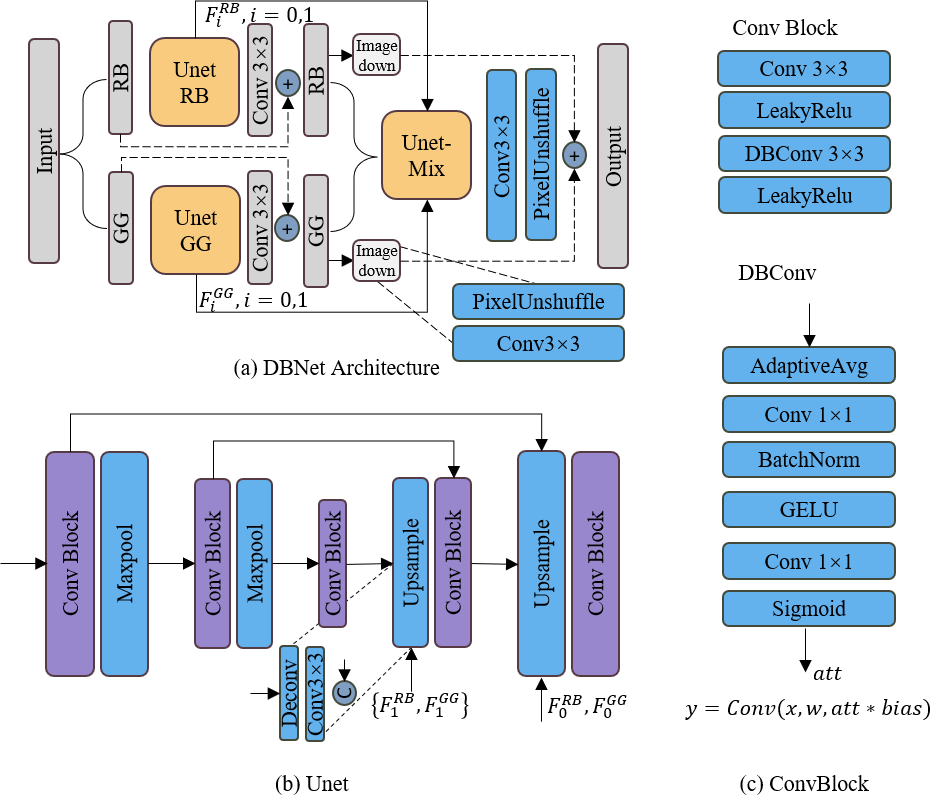}
    \caption{Team TongJi-IPOE. Overiew of the proposed DBNet.}
    \label{fig:TongJi-IPOE_diagram}
\end{figure}

Team \textbf{TongJi-IPOE} proposed a lightweight method for RAW image reconstruction from sRGB,  named DBNet, a dynamical bias convolution network.  The contributions of the proposed network are as follows: (i) a dynamical bias convolutional DBConv (see Fig. \ref{fig:TongJi-IPOE_diagram} c)was proposed to meet the needs of multiple data reconstruction and improve the fitting ability of the network,  (ii) a channel-mix processing network structure was proposed, which initializes the input sRGB image into RGGB channels, so that the design can simulate the RAW color pattern, as shown in Fig. \ref{fig:TongJi-IPOE_diagram} a. (iii) Pixel unshuffled and conv1×1 were used to downsample to avoid feature bias caused by interpolation downsampling.

\textbf{Network architecture} For RAW images, the green channel occupies half of the total pixels. Reference \cite{Liu2020joint} proposed SGNet for joint demosaicing and denoising tasks. For the task of reconstructing RAW from sRGB, there are differences in the difficulty of restoring different channels of R, G, and B. Inspired by \cite{Liu2020joint}, we proposed DBNet, as shown in Fig. \ref{fig:TongJi-IPOE_diagram} a. For the input sRGB image, we first initialize the image to RGGB color mode and divide it into two groups: RB and GG. We then input Unet RB and Unet GG for channel wise restoration. Further finetune through Unet Mix to obtain restored RAW images.  And image downsampling is only performed during output

\textbf{DBNet} The design of lightweight convolutional neural networks usually leads to the decline of model performance. To solve this problem, researchers improve the expression ability of models by establishing the relationship between input and convolution parameters, and adaptively adjusting convolution kernel parameters, such as CondConv\cite{RN183}, DyConv\cite{chen2020dynamic}, ODConv\cite{RN28} and DOConv\cite{RN16}. These methods will lead to a sharp increase in the number of parameters and increase the difficulty of training. Inspired by the need for registers in the visual transformer \cite{darcet2023vision}, we proposed that for visual tasks, the convolution layer needs a dynamic bias. In order to achieve the balance between lightweight design requirements and restoration performance, we propose a dynamic offset convolution DBConv. As shown in the Fig. \ref{fig:TongJi-IPOE_diagram} c, we dynamically adjust the bias parameters according to the input image to improve the expression ability of the model. Compared with the traditional convolution, the improvement of parameter and computation is almost negligible.

\textbf{Implementation Details} We implement our approach on single NVIDIA Geforce RTX 3090Ti GPU using the pytorch framework. We utilize AdamW optimize with ${{\beta }_{1}}=0.9$ and ${{\beta }_{2}}=0.999$ to optimize our proposed network. At the first stage, our model is trained for 92000 iterations, and the fixed learning rate is $ 1\times 10^{-3}$ . In the second stage, our model is trained for 208000 iterations, and the minimum learning rate is $1 \times 10^{-6}$, which is adjusted with the cosine annealing scheme. Finally, finetune 300000 iterations with a fixed learning rate of $1 \times 10^{-5}$.   Only the dataset provided was used in the training phase, and the patch size is set as $128 \times 128$.
\subsection{Lightweight U-Net for RGB to RAW Reconstruction}
\label{sec:unisoc}

\begin{center}

\vspace{2mm}
\noindent\emph{\textbf{Team Unisoc}}
\vspace{2mm}

\noindent\emph{Chen Wu~$^1$,
Zhansheng Wang~$^2$,
Pengbo Zhang,
Jiazi Huang}

\vspace{2mm}

\noindent\emph{Unisoc, China}

\vspace{2mm}


\end{center}


\paragraph{Method Description}

We present ULite\cite{10317244}, a U-Net\cite{DBLP:journals/corr/RonnebergerFB15}-based architecture specifically designed for efficient RGB to RAW image reconstruction. Our method focuses on parameter efficiency and computational performance while maintaining high-quality outputs.

ULite follows an encoder-decoder structure with several key innovations:

\begin{itemize}
\item \textbf{Cross-Domain Mapping:} Our architecture uniquely generates both a transformation 
matrix $M$ and an RGB domain image $I_{RGB}$. The final reconstructed RAW image is computed 
as $I^{'}_{RAW}=I_{RGB}*(M)^{-1}$, enabling more effective domain translation. 
At the same time, during the training phase, the $AWB$ and $CCM$ matrices can be extracted 
from the image metadata to obtain $M=AWB*CCM$. The network outputs $M$ and the final $I_{RAW}$.

\begin{table}
    \centering
    \begin{tabular}{l c c c}
         Method & PSNR & Params (M) & MACs (G)   \\
         \hline
         ULite\_S w/o M  & 32.72 & 0.19 & 2.32   \\
         ULite\_S w/  M  & 34.55 & 0.20 & 2.32   \\
         ULite\_L w/o M  & 35.02 & 2.45 & 16.04  \\
         ULite\_L w/  M  & 36.36 & 2.71 & 16.04  \\
    \end{tabular}
    \caption{Ablation study from Team Unisoc on the effect of color transformation matrix (M).}
\end{table}

\item \textbf{Efficient Architecture Design:} Our model employs Axial Depth-wise (AxialDW) 
convolutions that decompose operations into horizontal and vertical components, 
significantly reducing parameters while preserving spatial receptive field. 
For enhanced feature extraction, the bottleneck uses dilated AxialDW convolutions, 
while ULite\_L further incorporates Squeeze-and-Excitation blocks, 
Knowledge Bank Attention (KBA)\cite{Zhang2023kbnet}, and NAFBlocks\cite{chen2022simple} at strategic junctions.
\end{itemize}

\vspace{-2mm}

\paragraph{Dataset and Preprocessing}
We trained our models on the challenge dataset consisting of paired RGB-RAW images from iPhone-X and Samsung-S9 smartphones. The training data included: iPhone-X RGB-RAW pairs, Samsung-S9 RGB-RAW pairs, Additional low-quality (LQ) iPhone and Samsung data.

During training, we applied dynamic patch sizes (128×128 to 256×256) and used mask augmentation with a probability of 0.3, randomly masking regions of the image to improve robustness to incomplete data. No additional external datasets were used.

\vspace{-3mm}

\paragraph{Training Strategy.} We employed a multi-component loss function to optimize our models:

\begin{itemize}
\item \textbf{L1 Loss:} Primary loss component for pixel accuracy
\item \textbf{Color Loss:} A specialized loss that preserves color relationships between channels using both color ratio and difference constraints.
\item \textbf{Transformation Matrix Loss:} When valid metadata is available, 
we apply an MSE loss to the predicted transformation matrix $M^{'}$ compared to the ground truth 
matrix$M$, guiding the network to learn accurate domain transformations.
\item \textbf{Progressive Weighting:} Gradually increased the weight of color loss during training to stabilize convergence
\end{itemize}
Our loss function is formulated as:
\begin{align}
\mathcal{L}         &= \lambda_{L1} \cdot \mathcal{L}_{L1} + \lambda_{color} \cdot \mathcal{L}_{color} + \lambda_{M} \cdot \mathcal{L}_{M} \\
\mathcal{L}_{L1}    &= \frac{1}{N} \sum_{i=1}^{N} |I^{'}_{RAW} - I_{RAW}| \\
\mathcal{L}_{color} &= \frac{1}{N} \sum_{i=1}^{N} |\frac{I^{'}_{RAW}}{I_{RAW}} - 1| \\
\mathcal{L}_{M}     &= \frac{1}{N} \sum_{i=1}^{N} |M - M^{'}|
\end{align}
Where $\mathcal{L}$ is the total loss, $\mathcal{L}_{L1}$ is the pixel-wise L1 loss, 
$\mathcal{L}_{color}$ is the color consistency loss, 
and $\mathcal{L}_{M}$ is the transformation matrix loss. 
$\lambda_{L1}$, $\lambda_{color}$, 
and $\lambda_{M}$ are weighting factors set to 1.0, 0.001, 
and 0.1 respectively. 
$N$ represents the number of pixels in the image, 
$I^{'}_{RAW}$ is the predicted RAW image, 
$I_{RAW}$ is the ground truth RAW image, $M$ is the ground truth transformation matrix 
derived from metadata, and $M^{'}$ is the predicted transformation matrix. 
The color loss combines both ratio-based and 
difference-based penalties to ensure robust color reconstruction 
while avoiding numerical instabilities.

\begin{table}[t]
    \centering
    \begin{tabular}{l c c c}
         Method & PSNR & Params (M) & MACs (G)   \\
         \hline
         UNet            & 31.24 & 7.76 & 95.2   \\
         NAFNet          & 36.42 & 17.11& 64.29  \\
         ULite\_S (Ours) & 34.55 & 0.20 & 2.32   \\
         ULite\_L (Ours) & 36.36 & 2.71 & 16.04  \\
    \end{tabular}
    \caption{Team Unisoc ULite comparison with other methods. Our methods achieve strong PSNR scores while using fewer parameters and operations than competing methods.}
\end{table}

\begin{figure}[t]
    \centering
    \includegraphics[width=\columnwidth]{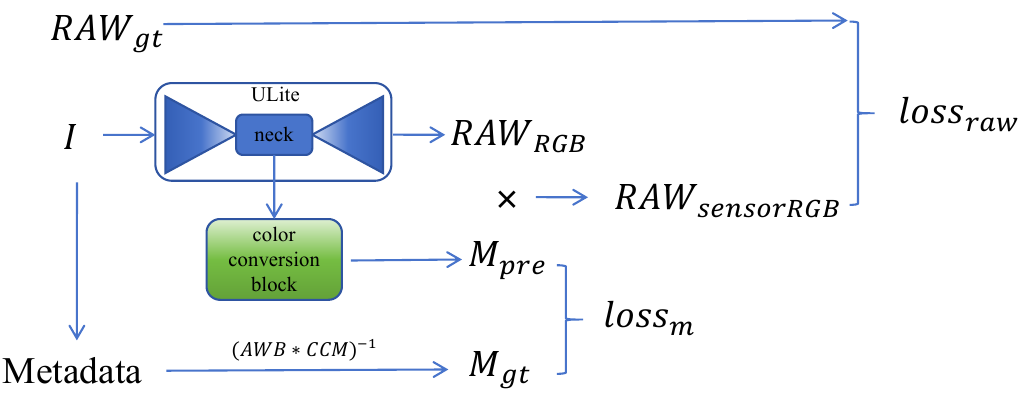}
    \caption{Overview of our ULite architecture proposed by team Unisoc. loss\_raw=$\mathcal{L}_{L1}+\mathcal{L}_{color}$, loss\_m=$\mathcal{L}_{M}$, loss\_total=loss\_raw+loss\_m}
    \label{fig:my_diagram}
\end{figure}

\vspace{-2mm}
\paragraph{Efficiency Analysis}
The efficiency stems from our use of separable convolutions, axial operations, and careful feature dimension management. Our models achieve favorable performance-to-parameter ratios compared to standard U-Net and NAFNet implementations. Our approach aligns with recent work demonstrating that lightweight CNN architectures can achieve competitive performance with significantly reduced parameters.

\vspace{-2mm}
\paragraph{Implementation Details}

We employed AdamW optimizer (initial lr=1e-4) with cosine annealing to 1e-7 over 500 epochs. 
Training ran for approximately 8 hours on an NVIDIA 4070super GPU with batch size 32. 
Our loss function combined L1 (weight 1.0) and color loss (weight 0.001), 
while data augmentation included random masking (prob 0.3, size 10-30\% of image) 
during training and 8-transformation TTA during inference.
\subsection{RAW Image Reconstruction Based on Global Appearance}
\label{sec:ivislab}

\begin{center}

\vspace{2mm}
\noindent\emph{\textbf{Team IVISLAB}}
\vspace{2mm}

\noindent\emph{Qinglin Liu~$^1$,
Wei Yu~$^2$,
Shengping Zhang~$^1$,
Xiangyang Ji~$^2$
}

\vspace{2mm}

\noindent\emph{$^1$ Harbin Institute of Technology, China\\
$^2$ Tsinghua University, China}

\vspace{2mm}


\end{center}


\begin{figure}[htbp]
    \centering
    \includegraphics[width=1.\linewidth]{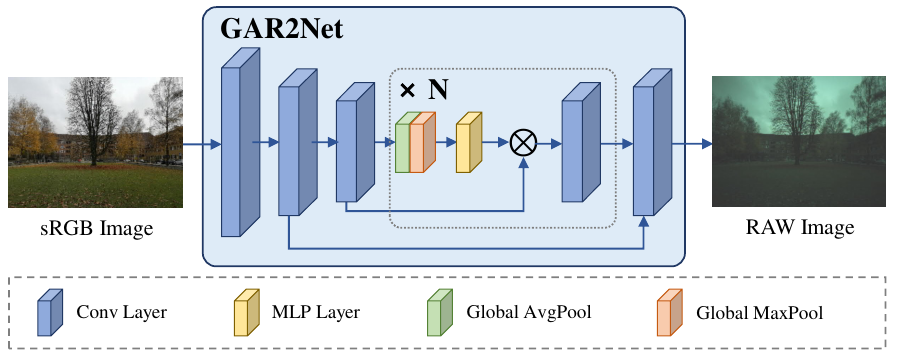}
    \caption{Architecture of GAR2Net by team IVISLAB.}
    \label{fig:garnet}
\end{figure}

\paragraph{Method Description}

To achieve RAW image reconstruction from sRGB images, we propose a Global Appearance-based RAW Reconstruction network (GAR2Net). The core idea is that the conversion from RAW to sRGB primarily involves local color transformations, along with global adjustments like white balance and exposure. Consequently, we focus on designing a network that effectively leverages global information, utilizing global average pooling and max pooling to build a Global Appearance Processing Module. Specifically, we introduce two variants of the network: a lightweight model and a full model. 

The GAR2Net network adopts an encoder-decoder architecture, as illustrated in Figure~\ref{fig:garnet}. At the beginning of the encoder, a series of convolutional layers are utilized to progressively downsample the input image while extracting rich local appearance features. To address the issue of gradient vanishing in deeper layers, we incorporate residual connections, which facilitate more stable and efficient training.
To capture global appearance information, we apply both global max pooling and global average pooling to the extracted local features. These pooled global descriptors are then passed through a multi-layer perceptron (MLP), and the resulting global context is used to modulate the local features through element-wise multiplication after a sigmoid activation. This mechanism allows the network to adaptively adjust local feature responses based on the overall image appearance. This process is performed repeatedly across multiple stages to gradually refine the feature representations.
In the decoder, we employ a series of upsampling convolutional blocks to progressively reconstruct the RAW image from the encoded features. PixelShuffle is used to increase spatial resolution efficiently while maintaining feature fidelity.

The GAR2Net framework consists of two variants: a full model and a lite model. The full model adopts a deeper and wider network backbone and incorporates channel attention modules to further enhance the integration of global contextual information. In contrast, the lite model utilizes a shallower and narrower architecture to reduce the number of parameters and computational cost, making it more suitable for resource-constrained environments.

\paragraph{Implementation Details}

GAR2Net is implemented using the PyTorch framework and trained on four NVIDIA RTX 3090 GPUs. During training, we use a batch size of 2, and input images are randomly cropped to a resolution of $384 \times 384$ pixels. The network is optimized for 2000 epochs using the AdamW optimizer, with an initial learning rate set to $4 \times 10^{-5}$. To ensure stable convergence and improved performance, a cosine annealing scheduler is employed to gradually decay the learning rate throughout training.
The overall loss function combines both $\ell_1$ and $\ell_2$ losses, with equal weighting coefficients $\lambda_1 = 1$ and $\lambda_2 = 1$ to balance pixel-wise accuracy and robustness. 

\noindent\textbf{Loss Function} To train GAR2Net, we define a reconstruction loss $\mathcal{L}_r$ for the estimated RAW image $I_{r}$ as follows:
\begin{equation}
\mathcal{L}_r = \lambda_1 \, \text{L}_1(I^{pred}, I^{gt}) + \lambda_2 \, \text{L}_2(I^{pred}, I^{gt})
\end{equation}
Here, $\lambda_1$ and $\lambda_2$ are coefficients that balance the loss terms. The function $\text{L}_1(\cdot,\cdot)$ represents the mean absolute error, while $\text{L}_2(\cdot,\cdot)$ denotes the mean squared error. The terms $I^{pred}$ and $I^{gt}$ refer to the predicted and ground truth images, respectively.

\subsection{Flexible Up/Down Sampling for ReverseISP using PixelShuffle/Unshuffle}
\label{sec:vip}

\begin{center}

\vspace{2mm}
\noindent\emph{\textbf{Team VIP}}
\vspace{2mm}

\noindent\emph{Minkyung Kim, Kyungsik Kim, Hwalmin Lee, and Jae-Young Sim}

\vspace{2mm}

\noindent\emph{Graduate School of Artificial Intelligence, Ulsan National Institute of Science and Technology, Republic of Korea}

\vspace{2mm}


\end{center}


\paragraph{Method Description}

\begin{figure*}[th]
	\centering
	\includegraphics[width=1\linewidth]{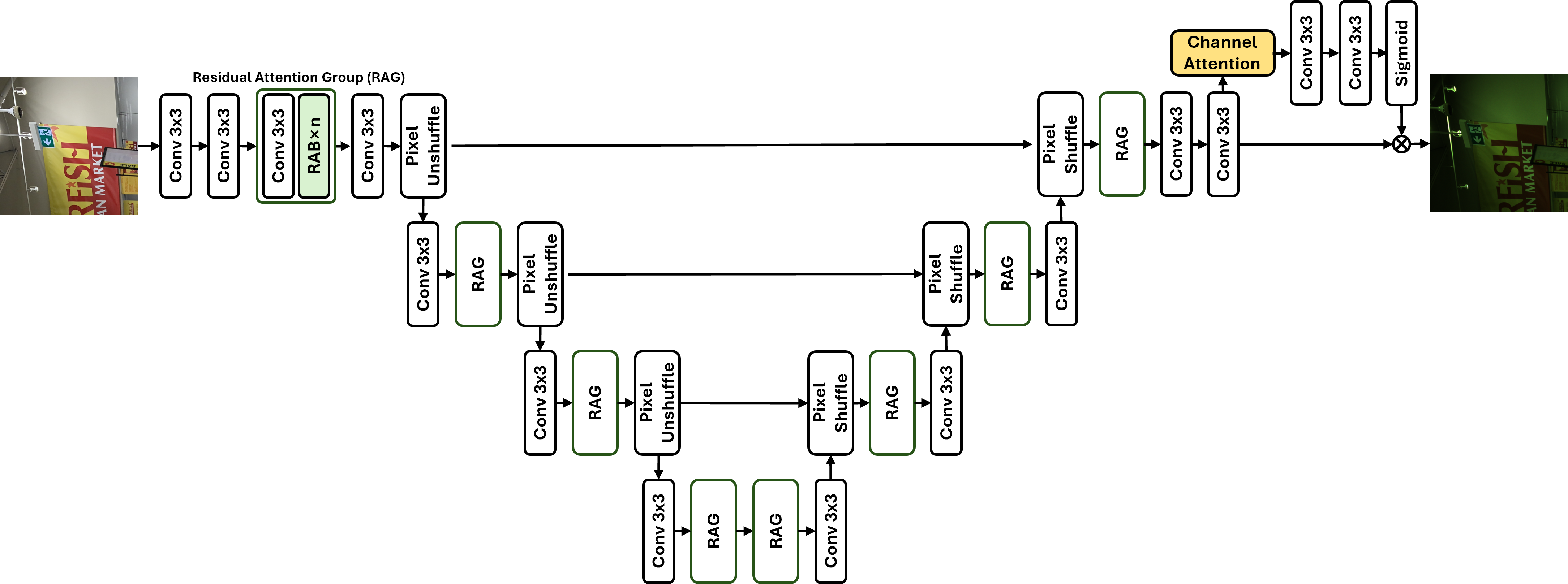} 
	\caption{Overall architecture of the proposed network for Reverse ISP by Team VIP.}
	\label{fig:overall}
\end{figure*}

\begin{figure}[t]
    \centering
    \includegraphics[width=1\linewidth]{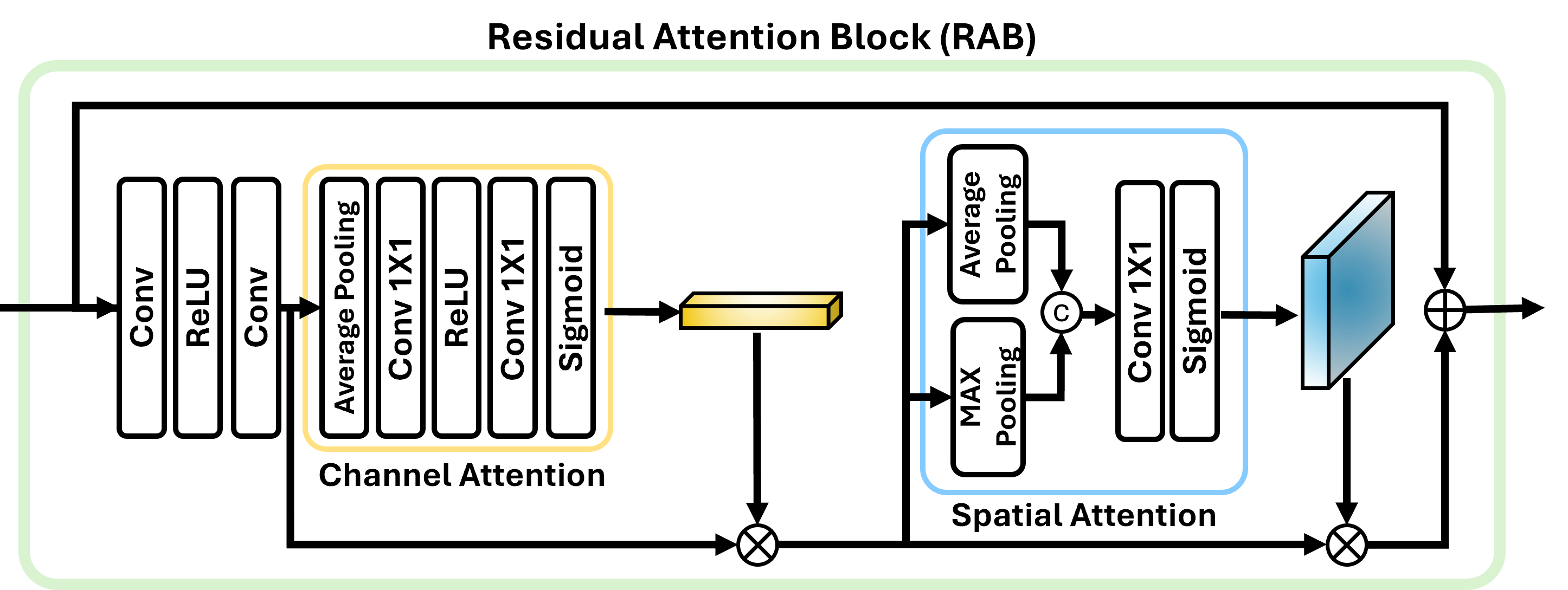} 
    \caption{Structure of the Residual Attention Block~\cite{conde2022reversed}.}
    \label{fig:attention_block}
\end{figure}

The overall architecture of the proposed model is shown in Figure~\ref{fig:overall}. The backbone follows the U-Net~\cite{ronneberger2015u} encoder-decoder structure with the skip connections. We incorporate the Residual Attention Groups~\cite{conde2022reversed} into each stage of the encoder and decoder. Each Residual Attention Group consists of several Residual Attention Blocks as illustrated in Figure~\ref{fig:attention_block}. 

The key modification in our architecture compared to the existing MiAlgo~\cite{conde2022reversed} is the replacement of the up/down sampling layers. The DWT layers used for downsampling and upsampling in MiAlgo are associated with the fixed filters, and thus pre-determined frequency subbands for decomposition/reconstruction may limit the flexibility in modeling the inverse ISP mappings. To enhance the flexibility of the network, we instead employ \texttt{nn.PixelUnshuffle} and \texttt{nn.PixelShuffle}~\cite{shi2016real} which yield learnable sampling integrated with the convolution operations. These convolutions learn task-specific feature aggregation (post-unshuffle) or preparation (pre-shuffle), bypassing the constraints of DWT's fixed frequency partitioning. This learned adaptivity is expected to improve the modeling of inverse ISP mappings.

\paragraph{Implementation Details}

We trained the model solely on the dataset provided in the challenge. From the official training set, 20\% was randomly set aside for a validation subset. No additional external datasets were used. 

\begin{itemize}
    \item \textbf{Optimizer and Learning Rate:} The AdamW optimizer~\cite{loshchilov2017decoupled} was employed with an initial learning rate of 1e-4. The cosine annealing learning rate scheduling~\cite{loshchilov2016sgdr} was used.
    \item \textbf{GPU:} Training was conducted on 2x NVIDIA A100 GPUs, each with 40GB of memory.
    \item \textbf{Datasets:} We utilized the official training dataset provided by the challenge organizers. Any pairs explicitly marked as Low-Quality (LQ) were excluded from our training dataset. 
    \item \textbf{Training Time:} Training was configured for 300,000 iterations using cosine annealing learning rate scheduling. We employed an early stopping strategy: training was stopped after 50,000 iterations (requiring 13 hours of training time on 2x NVIDIA A100 40GB) as we observed saturation in performance on the validation set. The checkpoint corresponding to the best validation score achieved within this period was used for the final evaluation reported in this paper.
    \item \textbf{Training Strategies:} The model was trained end-to-end using the L1 loss as the objective function. We used a batch size of 32, distributed across the two GPUs. 
    \item \textbf{Efficiency Optimization Strategies:} Beyond utilizing Automatic Mixed Precision (AMP) during training, no other optimization techniques (such as pruning or quantization) were employed.
\end{itemize}

\newpage
\subsection{ResUNet for RAW Image Reconstruction}
\label{sec:lvg}

\begin{center}

\vspace{2mm}
\noindent\emph{\textbf{Team  LVGroup-HFUT}}
\vspace{2mm}

\noindent\emph{Hekun Ma~$^1$,
Huan Zheng~$^2$,
Yanyan Wei~$^1$,
Zhao Zhang~$^1$, \\
$^1$ Hefei University of Technology, China\\
$^2$ University of Macau, China\\}

\vspace{2mm}

\end{center}


\paragraph{Method Description}

Our approach tackles the challenge of reconstructing RAW images from sRGB inputs using a deep learning framework. We employ a U-Net architecture~\cite{ronneberger2015u} with residual blocks to effectively capture and reconstruct the high-bit-depth details of RAW data from sRGB inputs. The model integrates encoder-decoder skip connections and leverages ensemble inference for improved performance.

We do not use \textbf{pre-trained or external methods/models}; The model was trained from scratch on the challenge dataset. The U-Net comprises three encoder blocks (EBlocks) with downsampling and three decoder blocks (DBlocks) with upsampling. Residual blocks enhance feature extraction and gradient flow. The input sRGB (3 channels) is mapped to RAW (4 channels) via convolutional layers, with skip connections aiding detail preservation. The model achieves low reconstruction losses (L1 and frequency-domain), Inference employed \textbf{ensemble techniques} using all model checkpoints, averaging predictions after horizontal and vertical flipping to enhance output quality. We used the dataset provided with the challenge \emph{ (sRGB-RAW pairs from devices such as the iPhone X and Samsung S9)}. The preprocessing included normalizing the images to [0,1] and converting them to tensors. No additional datasets were used. 

\begin{figure}[t]
    \centering
    \includegraphics[width=\linewidth]{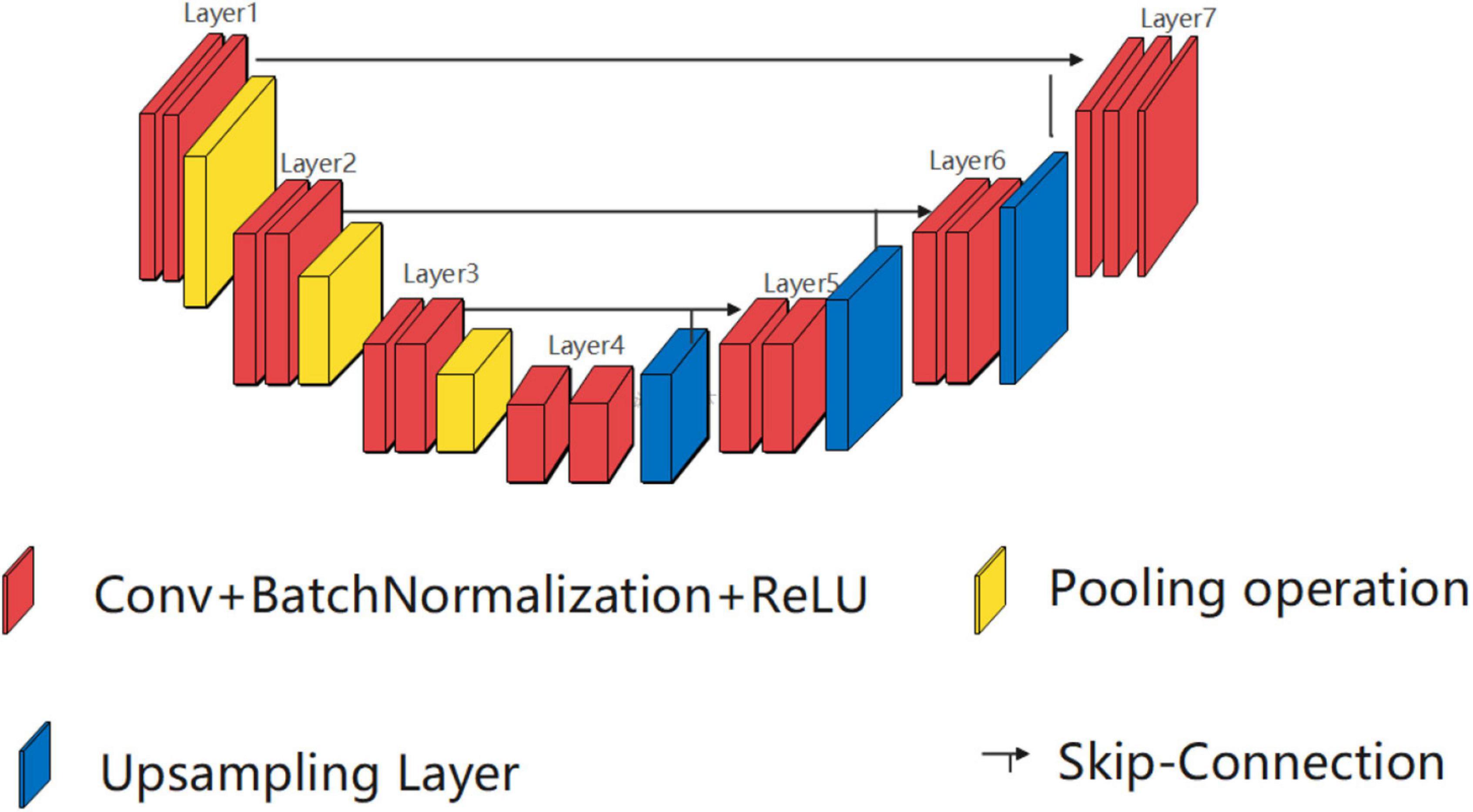} 
    \caption{The architecture of our ResUNet for RAW image reconstruction.}
    \label{fig:lvg}
\end{figure}

\paragraph{Implementation Details}

\begin{itemize}
    \item \textbf{Framework:}PyTorch
    \item \textbf{Optimizer and Learning Rate:}Training spanned 2000 epochs with a batch size of 4, using the Adam optimizer (lr=0.0002) and a multi-step learning rate scheduler (milestones at 1000, 1500, 1800, 2000; decay factor 0.5).
    \item \textbf{Efficiency:}Our general model has 4M parameters, trained on two RTX 4090 GPU for 30 hours with 2000 epochs. Inference runtime is optimized via PyTorch and ensemble averaging.
    \item \textbf{Datasets:} Challenge dataset with sRGB (1024x1024) and RAW (512x512) pairs. Augmentations included random cropping (768x768 for sRGB, 384x384 for RAW), horizontal, and vertical flipping.
    \item \textbf{Training Time:}Our general model has ~4M parameters and was trained on two RTX 4090 GPUs for ~30 hours over 2000 epochs. Inference runtime is optimized via PyTorch and ensemble averaging.
    \item \textbf{Training Strategies:} End-to-end training with combined L1 content loss and frequency reconstruction loss. Resume functionality was implemented for robustness.
    \item \textbf{Efficiency Optimization Strategies:} Residual blocks reduce computational overhead, while ensemble inference with all checkpoints balances accuracy and efficiency.
\end{itemize}

\subsection{NAFBlock-Enhanced UNet for Efficient RAW Image Reconstruction}
\label{sec:unafnet}

\begin{center}

\vspace{2mm}
\noindent\emph{\textbf{Team UNAFNet}}
\vspace{2mm}

\noindent\emph{Jing Fang~$^1$,
Meilin Gao~$^2$\\
Xiang Yu~$^3$\\
$^1$ School of Artificial Intelligence, Xidian University\\
$^2$ School of Artificial Intelligence, Xidian University\\
$^3$ School of Computer Science, Northeastern University\\}

\vspace{2mm}

\end{center}


\begin{table}[t]
    \centering
    \begin{tabular}{l c c}
         Method & PSNR & SSIM\\
         \hline
         Unet(baseline) & 23.18 & 0.67 \\
         RE-RAW & 27.28 & 0.76 \\
         RE-RAW+NAF & 29.13 & 0.84  \\
         RE-RAW+NAF+SSIM & 28.94 & 0.83  \\
         Unet+SSIM & 28.18 & 0.87  \\
         Unet+NAF+SSIM & 31.56 & 0.94  \\
         Unet+NAF+SSIM+Hard-Log-loss & 31.87 & 0.94  \\
    \end{tabular}
    \caption{Different experimental results by Team UNAFNet.}
    \label{tab:results}
\end{table}

\begin{figure}[t]
    \centering
    \begin{subfigure}[b]{0.22\textwidth}
        \includegraphics[width=\textwidth]{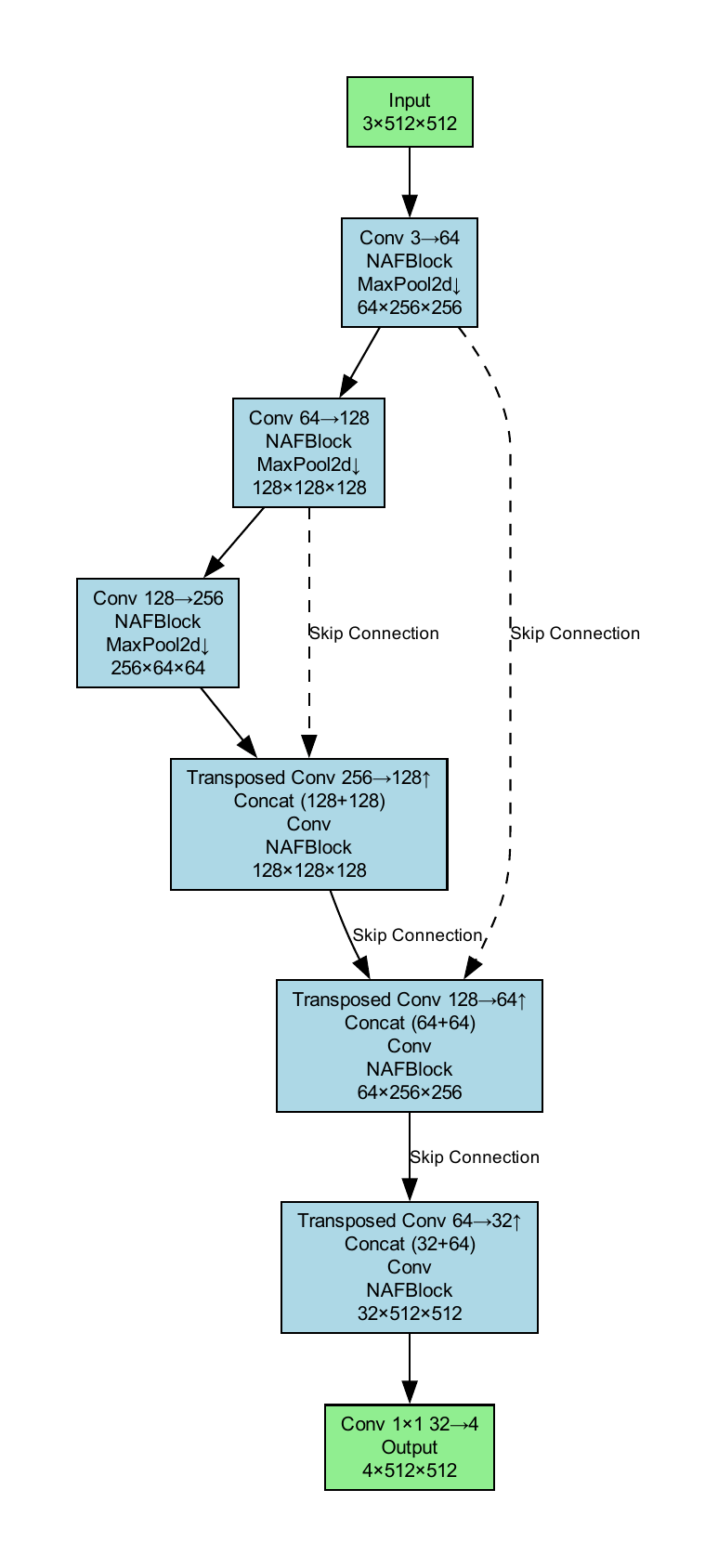} %
        \caption{UNAFNet}
        \label{fig:sub_a}
    \end{subfigure}
    \begin{subfigure}[b]{0.1\textwidth}
        \includegraphics[width=\textwidth]{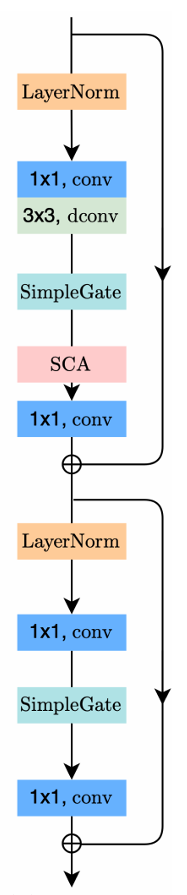}
        \caption{NAFBlock}
        \label{fig:sub_b}
    \end{subfigure}
    \caption{(a) Overview of our proposed network architecture. The model follows a UNet structure with NAF blocks and skip connections. The encoder path processes RGB input through three levels of feature extraction, while the decoder path reconstructs the RGGB RAW output.(b) Our proposed Nonlinear Activation Free
    Network’s block. It uses Simplified Channel Attention(SCA)
    and SimpleGate respectively.  }
    \label{fig:unafnet1}
\end{figure}

Our network architecture is inspired by the UNet structure but incorporates modern components for improved performance. As shown in Figure~\ref{fig:unafnet1}, The model consists of an encoder-decoder structure with skip connections, where each block is enhanced with  a nonlinear activation free block, noted as NAF block, which is depicted in Figure~\ref{fig:unafnet1}. The NAFBlock~\cite{chen2022simple} combines LayerNorm and SimpleGate mechanisms for better feature processing.

The encoder path processes the input RGB image through three levels of feature extraction, each containing:
\begin{itemize}
    \item A 3×3 convolutional layer for channel expansion
    \item A NAF block for feature processing
    \item A max pooling layer for spatial reduction
\end{itemize}

The decoder path symmetrically reconstructs the RAW image through:
\begin{itemize}
    \item Transposed convolution for spatial upsampling
    \item 3×3 convolution for feature processing
    \item NAF block for enhanced feature representation
    \item Skip connections from corresponding encoder levels
\end{itemize}

The final output head converts the features to RGGB format using a 1×1 convolution layer.

\begin{figure}[t]
    \centering
    \begin{subfigure}[b]{0.22\textwidth}
        \includegraphics[width=\textwidth]{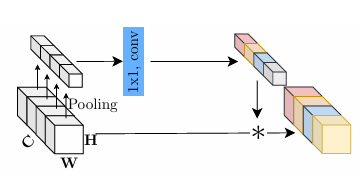} %
        \caption{SCA}
        \label{fig:sub_a}
    \end{subfigure}
    \begin{subfigure}[b]{0.22\textwidth}
        \includegraphics[width=\textwidth]{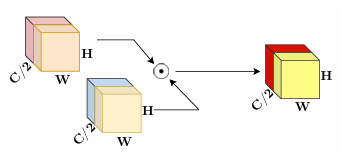}
        \caption{SG}
        \label{fig:sub_b}
    \end{subfigure}
    \caption{ (a) Simplified Channel Attention (SCA), and (b) Simple Gate (SG). \(\odot\)/\(\ast\): element-wise/channel-wise multiplication  }
    \label{fig:unafnet2}
\end{figure}

\paragraph{Training Strategy}

We train our model using a combination of three loss functions:
\begin{itemize}
    \item Mean Squared Error (MSE) loss for pixel-wise accuracy
    \item Structural Similarity Index Measure (SSIM) loss for perceptual quality
    \item Hard Log loss for better handling of extreme values and edge cases
\end{itemize}

The total loss is formulated as:
\[ L_{total} = L_{MSE} + 0.05 \times L_{SSIM} + 0.1 \times L_{hardlog} \]

where $L_{hardlog}$ is defined as:
\[ L_{hardlog} = -\mathbb{E}[\log(1 - \min(|x - y|, 1) + \epsilon)] \]

with $\epsilon = 10^{-6}$ to ensure numerical stability, and $x$ and $y$ representing the predicted and ground truth values respectively.

This combination allows us to optimize for both pixel-level accuracy (through MSE), structural and perceptual similarity (through SSIM), and robust handling of outliers (through hard log loss).

\paragraph{Implementation details}
\begin{itemize}
    \item \textbf{Framework:}PyTorch
    \item \textbf{Optimizer and Learning Rate:}Adam, The initial learning rate is\(10^{-4}\), and a dynamic learning rate scheduling strategy of cosine annealing that restarts every 16 epochs is adopted.
    \item \textbf{GPU:} NVIDIA A40 GPU.
    \item \textbf{Training Time:} 8 Hours.
    \item \textbf{Parameter Quantity:} 1669.89K.
\end{itemize}

\newpage
\begin{figure*}[t]
    \centering
    \begin{subfigure}[t]{0.6\textwidth}
        \centering
        \includegraphics[width=\linewidth]{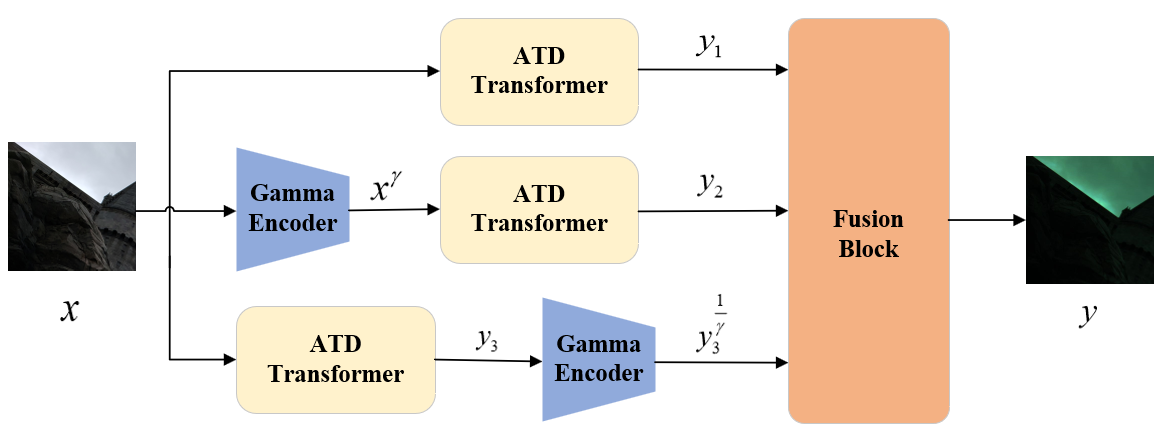}
        \caption{TDMFNet Architecture}
        \label{fig:sub1}
    \end{subfigure}
    \begin{subfigure}[t]{0.6\textwidth}
        \centering
        \includegraphics[width=\linewidth]{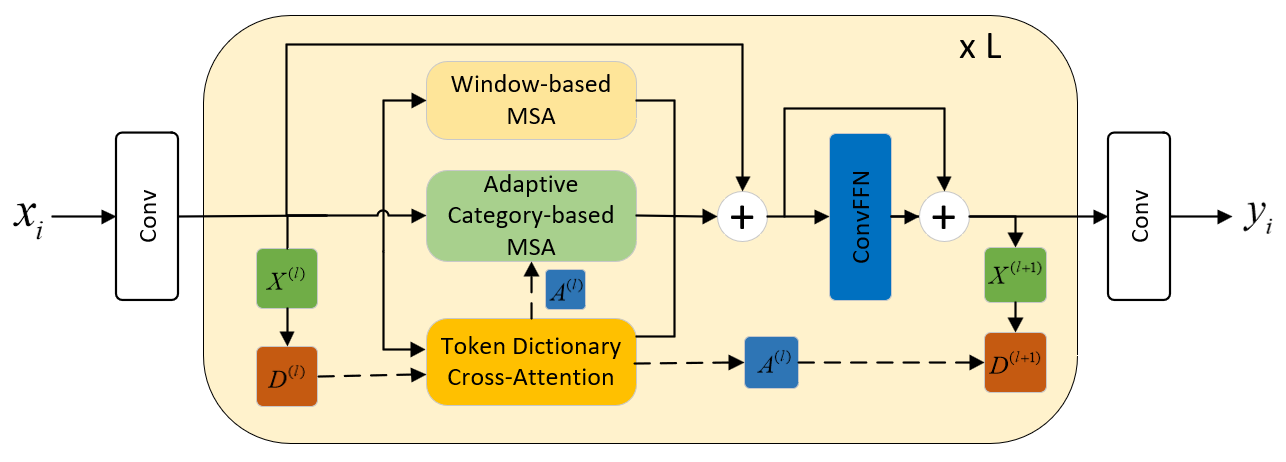}
        \caption{ATD Transformer Block\cite{zhang2024transcending}}
        \label{fig:sub2}
    \end{subfigure}
    \caption{Overview of the proposed TDMFNet.}
    \label{fig:TDMFNet}
\end{figure*}

\subsection{TDMFNet: Token Dictionary based Multi-path Fusion Network for sRGB-to-RAW Image Reconstruction}
\label{iirlab}

\begin{center}

\vspace{2mm}
\noindent\emph{\textbf{Team IIRLAB}}
\vspace{2mm}

\noindent\emph{Shangbin Xie~$^1$,
Mengyuan Sun~$^1$,
Huanjing Yue~$^1$,
Jingyu Yang$^{1}$ \\
$^1$ Tianjin University\\}

\vspace{2mm}

\end{center}


\paragraph{Method Description}

We propose a dual-stage framework named Token Dictionary based Multi-path Fusion Network for sRGB-to-RAW Image Reconstruction (TDMFNet) as illustrated in Fig~\ref{fig:TDMFNet}. It comprises a multi-path reconstruction network and an adaptive fusion network. In the first stage, we construct three mapping relationships and use parallel models to learn separately. The specific mapping relationships are defined as follows:  (a) \( \text{sRGB} \rightarrow \text{RAW} \), (b) \( G(\text{sRGB}) \rightarrow \text{RAW} \), (c) \( \text{sRGB} \rightarrow G^{-1}(\text{RAW}) \), where \( G \) represents the gamma scaling, and \( G^{-1} \) denotes its inverse process, which can be formulated as: 
\begin{equation}\label{newton2}
G(x) = x^\gamma
\end{equation}
For simplicity, the hyper-parameter \(\gamma\) is set to \(2.2\). As the three pathways exhibit distinct reconstruction performance under different scenarios, we introduce an adaptive fusion module to integrate their respective strengths. Specifically, the fusion module calculates weights \( w_{c, p} \), ensuring that the final output is \( y_{c} = \sum_{p=0}^{2} w_{c,p} \cdot x_{c,p} \), where \( c \) denotes the color channel, \( p \) represents the restoration path and \(x \) is the output of three paths. 
 
For each RAW reconstruction module, ATD\cite{zhang2024transcending} based network is employed. We introduce a group of adaptive token dictionary to learn RAW image priors from the training data. The dictionary is further used to classify image tokens and perform attention of tokens that belong to the same category. The category-based self-attention is performed between distant but similar tokens for enhancing input features, so that the receptive field is expanded to global image, which is well-suited for the sRGB-to-RAW conversion task.

To provide comprehensive supervision for training TDMFNet, the loss is calculated both in the individual paths and in the final fusion output.
\begin{equation}\label{newton2}
L_{paths} = l(y,\hat{y}_0) + l(y,\hat{y}_1) + l(G^{-1}(y),\hat{y}_2)
\end{equation}
\begin{equation}\label{newton2}
L_{fusion} = l(y,\hat{y})
\end{equation}
The overall loss function \(L\) is represented by
\begin{equation}\label{newton2}
L_{fusion} = L_{paths}+L_{fusion}
\end{equation}
where \( y \) represents the ground truth image, \(\hat{y}_i\) signifies the restored RAW image from path \( i \), and \(\hat{y}\) corresponds to the output of the fusion network. Furthermore, the perceptual loss \(L_p\) \cite{johnson2016perceptual} is also introduced to the loss function:
\begin{equation}\label{newton2}
l(y,\hat{y}) = L_1(y,\hat{y})+\lambda \cdot L_p(y,\hat{y})
\end{equation}
When calculating \(L_p\), we average the G1 and G2 channels of RAW images to match the input channel number of pre-trained model. Experimental results demonstrate that the perceptual loss effectively suppresses the lateral artifacts caused by the ATD grouping strategy and enhances the color accuracy of the reconstructed images.  The weight \(\lambda\) is set to 0.01 to balance the the L1 and perceptual loss.

\paragraph{Implementation Details}

We employed a three-stage progressive training strategy: starting with \(192\times192\) patches and a batch size of 6 for the first 30 epochs, then increasing to \(256\times256\) patches and a batch size of 3 until epoch 80, and finally using \(384\times384\) patches with a batch size of 1 until convergence at 120 epochs. The initial learning rate is \(1 \times 10^{-4}\) and changes with cosine annealing scheme to \(1 \times 10^{-6}\). The training utilizes the Adam optimizer with $\beta_{1,2}$ parameters \([0.9, 0.999]\). All experiments are conducted with the PyTorch framework on two NVIDIA GeForce RTX 4090D GPUs.

\subsection{Res-CSP Network}
\label{sec:changan}

\begin{center}

\vspace{2mm}
\noindent\emph{\textbf{Team Chang’an University}}
\vspace{2mm}

\author{Huize Cheng, Shaomeng Zhang, Zhaoyang Zhang, Haoxiang Liang \\
Chang’an University \\
}

\vspace{2mm}

\end{center}


\begin{figure*}[t]
    \centering
    \includegraphics[width=0.9\textwidth]{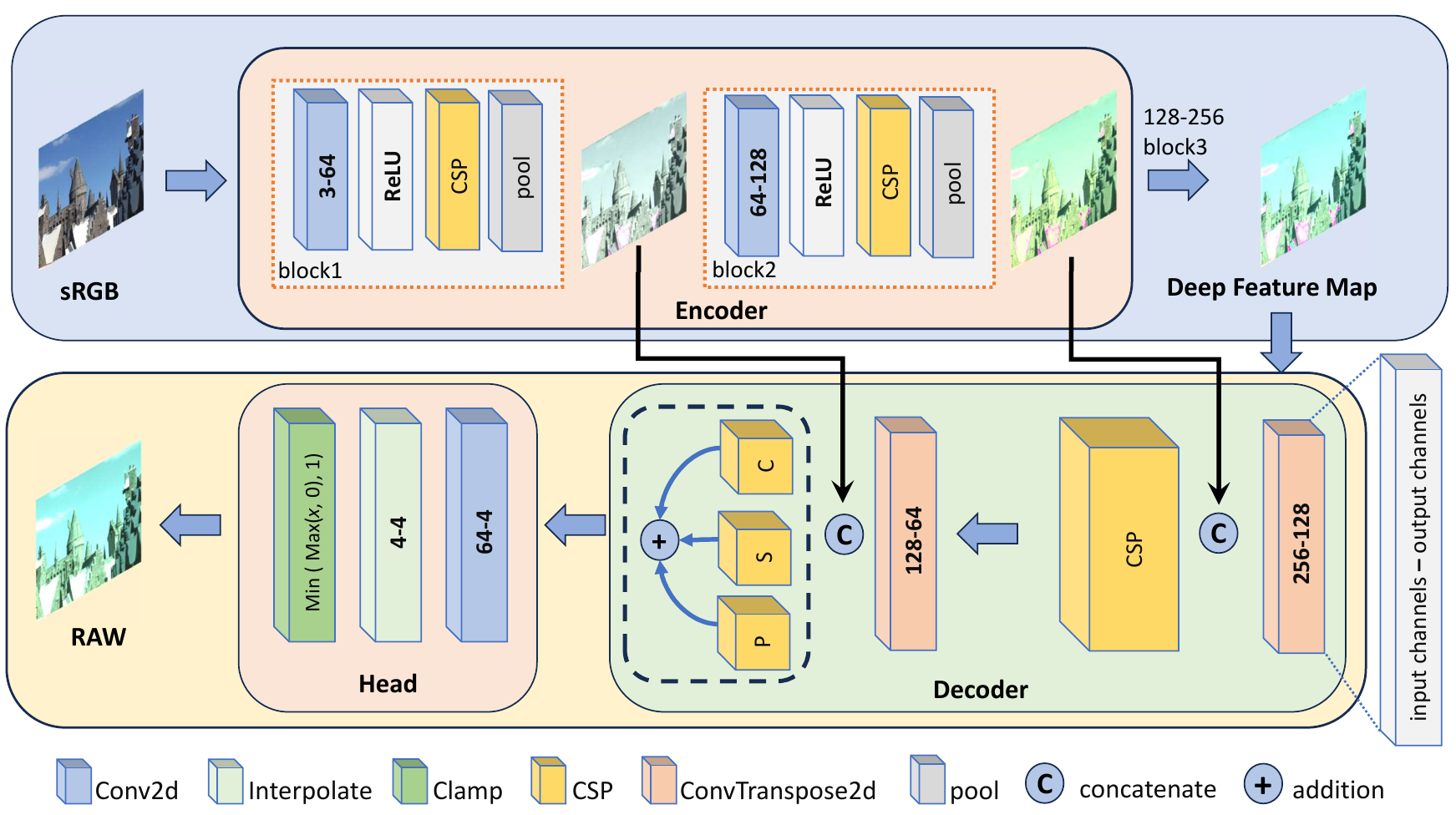}  
    \caption{
        \textbf{Res-CSP Network}: combines the benefits of ResNet with the feature extraction capabilities of the attention mechanism. 
    }   
    \label{fig:1}
\end{figure*}

The team proposes a Res CSP network based on residual connections and CSP modules for solving ISP reverse engineering problems and image super-resolution tasks. The experimental results show that the model achieved excellent performance of 29.78 dB PSNR and 0.92 SSIM in RAW inverse transform tasks (in the target devices), and can maintain stable reconstruction accuracy even in high noise environments.

\begin{table}[t]
  \centering
    \caption{Performance effects of different models on the RGB2RAW Target test set. The model Res-CSP outperforms the other models in both PSNR metrics and SSIM metrics.}
  \begin{tabular}{@{}lccc@{}}
    \toprule
    Method  & Year & PSNR$\uparrow$  & SSIM$\uparrow$ \\
    \midrule
    DeepLabV3Plus~\cite{liu2024image} & 2018 &  24.3121  & 0.85 \\
    ReRAW~\cite{berdan2025reraw} & 2025 &  24.4536  & 0.84 \\
    UNet++~\cite{bousias2020evaluation} & 2018 &  \underline{28.6390}  & 0.88 \\
    TransUNet~\cite{chen2024transunet} & 2024 &  28.0682  & \underline{0.90} \\
    Ours & 2025 & \textbf{29.7786}  & \textbf{0.92} \\
    \bottomrule
  \end{tabular}

  \label{tab:changan}
\end{table}

By using the Res CSP module, L1 hard logarithmic loss enhances the feature selection ability of the model, weakens unimportant features, and improves the interpretability of the model. Our model achieves \textbf{29.7786}dB The PSNR. As for SSIM, our model achieves \textbf{0.92}, indicating that the reconstructed image is highly consistent with the original image in terms of structure and texture details.

\paragraph{Method Description}

Res-CSP Network designed for image processing tasks. The encoder, which plays a crucial role in feature extraction, is composed of multiple blocks that leverage convolutional layers, ReLU activations, and CSP modules to efficiently process input images.The core network structure is shown in Fig. \ref{fig:1}.\textbf{Encoder}: Responsible for extracting deep features from the input image. \textbf{Decoder}: Converts the deep feature map extracted by Encoder back to an output of the same size as the input image. \textbf{Head}: Converts the multi-channel feature maps from Decoder output to the final 4-channel RAW image.In this study, a new loss function, the L1 hard logarithmic loss, is proposed, which combines the properties of the L1 loss and the hard logarithmic loss.

\paragraph{Implementation Details}

\textbf{rgb2raw Dataset}~ The images were captured with three different smartphone cameras across diverse scenes and lighting conditions. This dataset comprises real noisy images and their corresponding ground truth, offering synchronized RAW domain sensor data (raw RGB) and sRGB-color space data. With 2952 ultra-high-resolution image pairs for model training and a validation set of 120 image pairs, the dataset provides a robust foundation for robust model development.

\textbf{Data Cleaning.}~ In this study, in order to improve the quality of blurred and noisy images in the dataset, \textbf{DnCNN} (Deep Neural Network for Image Denoising) is used for data preprocessing.DnCNN is a deep learning-based image denoising network that can effectively remove noise from an image while retaining the details and structural information of the image. Using the powerful denoising capability of DnCNN, we can pre-process images in the data set to improve image quality and provide higher quality input data for subsequent image analysis and processing tasks.

\textbf{Optimizer and Learning Rate.}~ Optimizer using Adam optimizer.The initial learning rate is set to 5e-5.The learning rate decay rate is set to 2e-6.The input of the Res CSP model is RGB images of (256, 256, 3) size. During the training process, an end-to-end training approach was adopted, with a total training time of approximately 33 hours. No additional data augmentation techniques were used during the training process.The total number of parameters for this model is approximately 4.06 million, and it was trained on NVIDIA V100 GPU.

\begin{table*}[t]
    \scriptsize
    \centering
    \resizebox{\textwidth}{!}{
    \begin{tabular}{r|c|c|c|c|c|c|c|c}
        
        \toprule
        Method & Input & Training Time & Train E2E & Extra Data & \# Params. (M) & FLOPs (G) & GPU  \\
        \midrule
        
         DualRAW~\ref{sec:dualraw} & $1024 \times 1024 \times 3$ & 40h & Yes & No & 1.6 & - & Nvidia H100 (80G) \\
         
         GAR2Net-Full~\ref{sec:ivislab}  & $256 \times 256 \times 3$ & About 120h & Yes & No & 4.496 & 17.97 & RTX3090 \\
         
         GAR2Net-Lite~\ref{sec:ivislab}  & $256 \times 256 \times 3$ & About 50h & Yes & No & 0.194 & 1.10 & RTX3090 \\

         DBNet~\ref{sec:tongji} & $128 \times 128 \times 3$ & 24h & Yes & No & 0.39  & - & 3090Ti \\
         
         VIP~\ref{sec:vip} & $1024 \times 1024 \times 3$ & 13h & Yes & No & 4.5 & - & A100 \\

         ResUnet~\ref{sec:lvg} & $768 \times 768 \times 3$ & 30h & Yes & No & 5 & - & RTX 4090 \\

         TDMFNet~\ref{iirlab} & $384 \times 384 \times 3$ & 24h & Yes & No & 2.37 & 78.11 & 4090D \(\times2\) \\

         \bottomrule
         
    \end{tabular}
    }
    \caption{Technical summary of the proposed solutions.}
    \label{tab:summary}
\end{table*}




\newpage

{\small
\bibliographystyle{ieeenat_fullname}
\bibliography{refs}
}

\end{document}